\documentclass[conference]{IEEEtran}
\IEEEoverridecommandlockouts

\usepackage[dvipsnames,table,xcdraw]{xcolor}
\usepackage[pagebackref=true,breaklinks=true,colorlinks,citecolor=ForestGreen]{hyperref}
\usepackage{ragged2e}

\usepackage{cite}
\usepackage{amsmath,amssymb,amsfonts}
\usepackage{algorithmic}
\usepackage{graphicx}
\usepackage{textcomp}

\usepackage{algorithm}
\usepackage{booktabs}
\usepackage{colortbl}
\usepackage{multirow}
\usepackage{pifont}
\usepackage[switch]{lineno}
\usepackage{ragged2e}
\usepackage{tabularx}
\usepackage{enumitem}

\def\BibTeX{{\rm B\kern-.05em{\sc i\kern-.025em b}\kern-.08em
    T\kern-.1667em\lower.7ex\hbox{E}\kern-.125emX}}
\begin{document}

\title{Controllable Expressive 3D Facial Animation via Diffusion in a Unified Multimodal Space
\thanks{$^{\star}$ Corresponding authors.}
\thanks{This work was supported by NSFC under 62272456.}
\thanks{We extend our special thanks to Ruoyu Chen for helping revise Figures 1, the Introduction, and the Abstract.}
}

\author{
\IEEEauthorblockN{Kangwei Liu$^{1,2}$, Junwu Liu$^{1,2}$, Xiaowei Yi$^{1,2}$, Jinlin Guo$^{3}$, Yun Cao$^{1,2,\star}$  }
\IEEEauthorblockA{$^1$ Institute of Information Engineering, Chinese Academy of Sciences, Beijing 100085, China}
\IEEEauthorblockA{$^2$ School of Cyber Security, University of Chinese Academy of Sciences, Beijing 100049, China} 
\IEEEauthorblockA{$^3$ Laboratory for Big Data and Decision, School of Systems Engineering, National University of Defense Technology} 
\IEEEauthorblockA{\{liukangwei, liujunwu, yixiaowei, caoyun\}@iie.ac.cn, gjlin99@nudt.edu.cn}
}

\maketitle

\begin{abstract}

Audio-driven emotional 3D facial animation encounters two significant challenges: (1) reliance on single-modal control signals (videos, text, or emotion labels) without leveraging their complementary strengths for comprehensive emotion manipulation, and (2) deterministic regression-based mapping that constrains the stochastic nature of emotional expressions and non-verbal behaviors, limiting the expressiveness of synthesized animations. To address these challenges, we present a diffusion-based framework for controllable expressive 3D facial animation. Our approach introduces two key innovations: (1) a FLAME-centered multimodal emotion binding strategy that aligns diverse modalities (text, audio, and emotion labels) through contrastive learning, enabling flexible emotion control from multiple signal sources, and (2) an attention-based latent diffusion model with content-aware attention and emotion-guided layers, which enriches motion diversity while maintaining temporal coherence and natural facial dynamics. Extensive experiments demonstrate that our method outperforms existing approaches across most metrics, achieving a 21.6\% improvement in emotion similarity while preserving physiologically plausible facial dynamics. Project Page: \url{https://kangweiiliu.github.io/Control_3D_Animation}.

\end{abstract}


\section{Introduction}
Recent advancements in audio-driven 3D facial animation~\cite{vocaset, meshtalk, faceformer, codetalker, emote, emotalk, imitator} have significantly enhanced realistic virtual characters in virtual reality, digital entertainment, and human-computer interaction. However, achieving expressive 3D facial animation remains challenging, particularly in effectively controlling emotions through multiple modalities \cite{emotalk}.

Current approaches to emotionally expressive 3D facial animation face two critical limitations. First, existing methods predominantly rely on single-modal control signals~\cite{emotalk, mead, styletalk, emote}, where videos excel at capturing temporal dynamics and subtle expressions, text enables abstract semantic-level control, and emotion labels provide efficient categorical guidance. However, these methods lack a unified framework to leverage such complementary strengths for comprehensive emotion manipulation. Second, current emotion control 3D facial animation models \cite{meshtalk, faceformer, emote} employ deterministic regression-based mappings that enforce a one-to-one correspondence between speech and facial motion, fundamentally limiting their ability to capture the stochastic nature of emotional expressions and non-verbal behaviors, resulting in reduced expressiveness in the synthesized animations.

\begin{figure}[!t]
  \centering
  \includegraphics[width=0.48\textwidth]{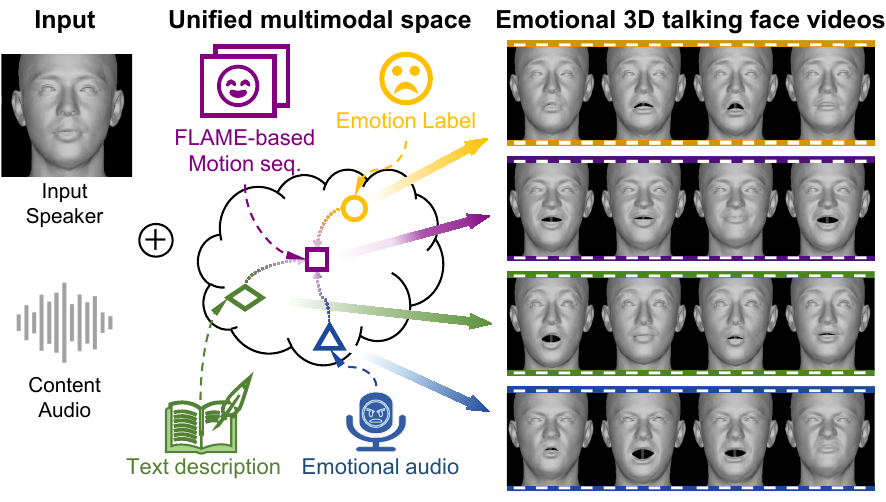}
  \vspace{-2mm}
  \caption{Illustration of our proposed method. Given an input speaker, a content audio clip, and an emotional control signal (FLAME-based motion sequence, text, label, or audio), our method generates stylized talking faces where the speaker articulates the content in the emotional style.}
  \vspace{-2mm}
  \label{fig:abstrac_fig}
\end{figure}


To address these challenges, we present a diffusion-based framework leveraging FLAME \cite{flame} for synthesizing expressive 3D facial animations, achieving both flexible emotion manipulation through a unified multimodal space (Fig. \ref{fig:abstrac_fig}) and diverse motion synthesis via latent diffusion modeling. Our framework introduces two key technical innovations: (1) A FLAME-centered multimodal emotion binding strategy that projects emotional representations from different modalities (text, audio, and emotion labels) into a unified latent space through contrastive learning, where FLAME serves as the central representation due to its rich semantic information derived from visual modality, enabling flexible emotion manipulation by leveraging their complementary characteristics. (2) An attention-based latent diffusion architecture that captures the inherent stochasticity of emotional expressions by incorporating learned noise patterns into the speech-to-motion mapping process. This architecture, enhanced with cross-modal attention mechanisms and emotion-conditioned layers, facilitates the synthesis of diverse non-verbal behaviors while preserving speech synchronization and emotional coherence.

\begin{figure*}[!t]
  \centering
  \includegraphics[width=0.95\textwidth]{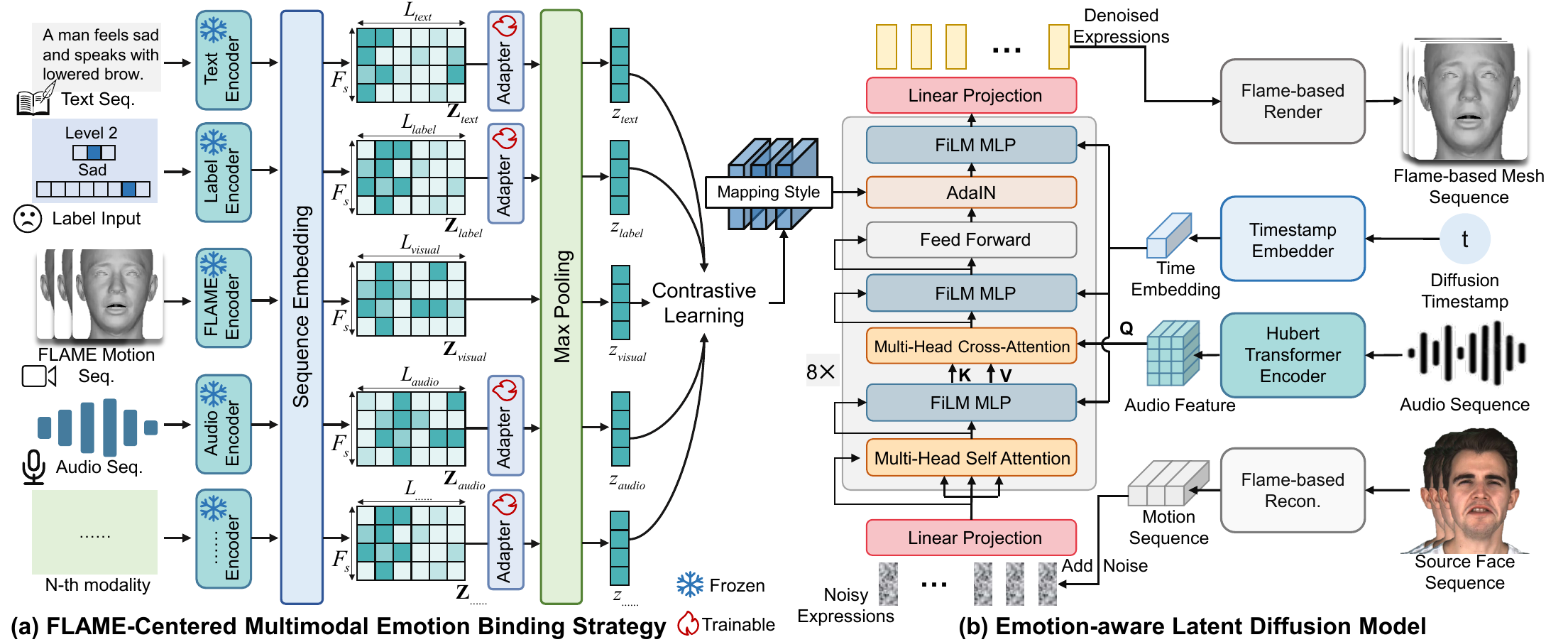}
  \vspace{-2mm}
  \caption{Overview of our proposed framework. Our approach unifies diverse modalities through (a) FLAME-centered multimodal emotion binding strategy and (b) an attention-based latent diffusion model that generates facial dynamics while maintaining speech synchronization and emotional consistency.}
  \label{fig:overview}
  \vspace{-2mm}
\end{figure*}

Extensive experiments validate the effectiveness of our framework. Our multimodal binding strategy, combined with a diffusion-based approach, demonstrates superior performance across different input modalities,  achieving a 21.6\% improvement in emotion similarity and enhancing both facial expression diversity and lip synchronization throughout the animations. 

The key contributions of our work are as follows: (1) A novel diffusion-based framework for expressive 3D facial animation that jointly addresses the limitations of single-modal control and deterministic generation through FLAME representation. (2) A multimodal emotion binding strategy that unifies emotional representations from different modalities through contrastive learning, enabling flexible control from different modality inputs. (3) An attention-based latent diffusion model that introduces controlled stochasticity while preserving speech synchronization, enhancing the diversity of emotional expressions. (4) Extensive experiments demonstrating the superiority of our method over state-of-the-art methods.





\section{Related Work}

\subsection{Speech-driven 3D Facial Animation}
Speech-driven facial animation has evolved from rule-based approaches \cite{avsp} to learning-based methods. Recent works leverage acoustic features or pre-trained speech embeddings \cite{wav2vec2, hubert} to drive various 3D representations, including morphable models \cite{sadtalker,talkface} and direct meshes \cite{vocaset, meshtalk, faceformer,codetalker,imitator}. While these deterministic methods achieve precise lip synchronization \cite{produb}, they often produce over-smoothed animations lacking expressiveness. Recent diffusion-based approaches \cite{facediffuser,diffusionmotion} show improved motion diversity but lack explicit emotion control mechanisms.

\subsection{Emotional Expression Control and Datasets}
Emotional expression control in facial animation primarily follows two paradigms. Label-based approaches \cite{emote, mead} rely on discrete emotion categories or continuous valence-arousal values, offering direct but limited control over expressions. Example-based methods \cite{imitator,emotalk,styletalk} enable more nuanced emotion transfer by learning from reference clips, but often struggle with temporal consistency and require complex optimization procedures. Recent attempts at multi-modal emotion control either handle each modality separately or employ simple feature concatenation, leading to inconsistent emotional representations across modalities.

The development of emotion control methods is closely tied to available datasets. While traditional datasets (BIWI \cite{biwi}, VOCASET \cite{vocaset}, Multiface \cite{multiface}) focus on neutral speech, specialized emotional datasets like MEAD \cite{mead} and RAVDESS \cite{RAVDESS} provide rich emotional annotations. Recent advances in FLAME-based \cite{flame} monocular reconstruction \cite{emoca} have enabled high-quality 3D face capture, bridging the gap between 2D emotional datasets and 3D animation requirements.

\section{Method}
We propose a diffusion-based framework for generating expressive 3D facial animations within the FLAME expression space, as illustrated in Fig. \ref{fig:overview}. Our framework consists of two key stages: (1) a FLAME-centered multimodal emotion binding strategy (Sec. \ref{sec:a}) that unifies emotional representations from different modalities, and (2) an attention-based latent diffusion model (Sec. \ref{sec:b}) that synthesizes diverse facial motions guided by the unified emotional representations.

\subsection{FLAME-Centered Multimodal Emotion Binding Strategy} \label{sec:a}
Existing methods for emotional facial animation rely on rigid single-modal control signals \cite{emotalk, mead, styletalk}, requiring separate models for different emotional inputs. To address this limitation, we propose a FLAME-centered multimodal emotion binding strategy that maps diverse emotional representations (text, audio, and emotion labels) into a unified expression space through contrastive learning, enabling flexible emotion manipulation from multiple modalities.

Specifically, we design modality-specific encoders to extract emotionally relevant features and align them into the FLAME expression space. Our framework incorporates four specialized encoders\footnote{The detailed implementation is provided in the supplementary material.}:

\noindent\textbf{FLAME Encoder.} To capture emotion-related visual features, we employ a transformer-based encoder \cite{attention} $\mathcal{E}_{\text{V}}$, pre-trained on the MEAD dataset through an emotion classification task. 

\noindent\textbf{Audio Encoder.} For emotional speech features, we utilize a pre-trained Wav2vec 2.0 model \cite{wav2vec2} $\mathcal{E}_{\text{A}}$, fine-tuned on speech emotion classification tasks.

\noindent\textbf{Text Encoder.} To process textual emotion descriptions, we employ a RoBERTa-based encoder \cite{roberta} $\mathcal{E}_{\text{T}}$.

\noindent\textbf{Label Encoder.} We implement a learnable embedding layer $\mathcal{E}_{\text{L}}$ for mapping discrete emotion categories into FLAME-aligned representations.

The encoding process for each modality can be formally expressed as:
\begin{equation}
    \mathbf{Z}_m = \mathcal{E}_m(\mathbf{IN}_m),~~~~m \in \{\text{V}, \text{A}, \text{T}, \text{L}, \text{\textit{etc.}}\},
\end{equation}
where $\text{V}$, $\text{A}$, $\text{T}$, $\text{L}$, $\text{\textit{etc.}}$ represent the visual, audio, text, label, and other modalities respectively, $\mathbf{IN}_m$ represents the input from modality $m$, $\mathcal{E}_m$ denotes the corresponding encoder, and $\mathbf{Z}_m \in \mathbb{R}^{L_m \times F_s}$ is the output of the corresponding encoders. The feature dimension is $F_s$, while the length is $L_m$, which can be different.

To address the temporal misalignment between modalities noted in \cite{seeg}, which can confuse the encoder if temporal correspondence is rigidly enforced during training, we aggregate emotion-relevant information in each feature sequence via the MaxPool operation:
\begin{equation}
\mathbf{z}_{m} = \text{MaxPool}(\mathbf{Z}_m),~~~~m \in \{\text{V}, \text{A}, \text{T}, \text{L}\},
\end{equation}
where $\mathbf{z}_m \in \mathbb{R}^{F_s}$ represents the MaxPool output for modality $m$. This operation highlights salient emotional features, ensuring effective contribution from all modalities to the FLAME-aligned embedding.

\noindent\textbf{Contrastive Learning for Multimodal Alignment.} 
To align representations from different modalities within the FLAME expression space, we employ a contrastive learning framework. For each modality, we introduce a lightweight adapter network (three fully connected layers) while keeping pre-trained encoder parameters fixed. The contrastive loss is defined as:
\begin{equation}
    \mathcal{L}_{\text{info}} = -\sum_{(i,j) \in \mathcal{P}} \log \frac{\exp(\text{sim}(\mathbf{z}_i, \mathbf{z}_j)/\tau)}{\sum_{k \in \mathcal{N}_i}\exp(\text{sim}(\mathbf{z}_i, \mathbf{z}_k)/\tau)} ,
\end{equation}
where $\mathcal{P}$ denotes positive pairs with matching emotional content, $\mathcal{N}_i$ represents negative samples for anchor $i$, $\text{sim}(\cdot,\cdot)$ computes cosine similarity, and $\tau$ is a temperature parameter. This objective ensures emotionally consistent representations are aligned in the FLAME expression space while separating those with different emotional content. These aligned representations serve as conditional signals to guide the subsequent facial motion synthesis process through our attention-based latent diffusion model.

\subsection{Emotion-aware Latent Diffusion Model} \label{sec:b}
To address the limited diversity in facial motion synthesis caused by deterministic speech-to-motion mapping, we propose an attention-based latent diffusion model that introduces stochasticity into the generation process while maintaining both speech synchronization and emotional consistency.

\noindent \textbf{Audio Encoding.} For robust speech content extraction, we leverage a pre-trained Hubert model \cite{hubert} fine-tuned on ASR tasks, which effectively isolates speech content from other factors (e.g., emotion, background noise).

Our generator employs a FLAME-space latent diffusion model. The forward process gradually diffuses Gaussian noise over \(T\) timesteps, mapping the initial 3D facial sequence \(\mathbf{Z}_0\) to noise-corrupted versions \(\mathbf{Z}_t\) through:
\begin{equation}
q(\mathbf{Z}_t \mid \mathbf{Z}_{0}) = \mathcal{N}(\sqrt{1 - \beta_t} \mathbf{Z}_{0}, \beta_t \mathbf{I}),
\end{equation}
where \(\beta_t\) determines the noise schedule. The reverse process reconstructs the original motion \(\mathbf{Z}_0\) by iteratively denoising from \(\mathbf{Z}_N \sim \mathcal{N}(0, \mathbf{I})\). To integrate the timestamp into our model, we use a feature-wise linear modulation (FiLM) layer \cite{film}. FiLM layers allow the network to adaptively adjust its computations, enhancing its ability to handle diverse tasks and input variations.

\noindent \textbf{Content-aware Style Integration.} To ensure both diverse facial dynamics and precise lip synchronization, we introduce two key components:

1) Content-aware attention layer: Traditional cross-attention uses an upper triangular mask, causing lip-sync issues due to its allowance for current queries to attend to past keys and values. To resolve this, we implement a diagonal masking strategy to ensure precise frame-level synchronization:
\begin{equation} 
\text{Attention}(\mathbf{Q}, \mathbf{K}, \mathbf{V}, \mathbf{M}) = \text{softmax}\left(\frac{\mathbf{Q} \mathbf{K}^T}{\sqrt{d}} + \mathbf{M}\right) \mathbf{V},
\end{equation}
where \(\mathbf{M}\) allows each query to attend only to its corresponding key and value, maintaining accurate lip-sync with audio input.

2) Emotion-guided AdaIN layer: We implement an adaptive instance normalization layer \cite{adain} that processes multimodal emotional style embeddings through a compact 3-layer MLP, transforming style vector \(\mathbf{z}_s\) into modulation parameters for flexible emotion control.

\noindent \textbf{Enhanced Training Strategy.} To achieve optimal emotion transfer, we adopt a classifier-free guidance approach \cite{classfree}. During training, conditional inputs (audio $\mathbf{A}$ and emotional style prompts $\mathbf{P}$) are randomly masked with 0.1 probability. At inference, emotional intensity is modulated via:
\begin{equation} 
\epsilon^{*}_{n} = s \epsilon_{\theta}(\mathbf{Z}_n, n, \mathbf{A}, \mathbf{T}, \mathbf{P}) + (1 - s) \epsilon_{\theta}(\mathbf{Z}_n, n, \varnothing, \mathbf{T}, \varnothing),
\end{equation}
where \(s\) controls the style-fidelity trade-off.

To ensure both accurate lip synchronization and faithful emotional consistency, we optimize three complementary losses:

1) \textbf{Lip Synchronization Loss:} Inspired by syncnet \cite{syncnet}, We design a FLAME-based lip-sync expert that evaluates the temporal correspondence between generated FLAME expressions and input audio features through contrastive learning\footnote{The detail of the FLAME-based lip-sync expert is shown in the supplementary material.}:
\begin{equation}
\mathcal{L}_{\text{sync}} = -\log \frac{\exp(\text{sim}(\mathbf{f}_e, \mathbf{f}_a)/\tau)}{\sum_{k}\exp(\text{sim}(\mathbf{f}_e, \mathbf{f}_a^k)/\tau)}, 
\end{equation}
where $\mathbf{f}_e$ and $\mathbf{f}_a$ denote the FLAME expression parameters and audio features respectively, and $\mathbf{f}_a^k$ represents negative audio samples.

2) \textbf{Emotion Consistency Loss:} To maintain emotional fidelity, we employ a FLAME-based emotion encoder (same as the one in the emotion alignment module) that measures the L2 distance between the generated expressions and target emotion embeddings:
\begin{equation}
\mathcal{L}_{\text{emo}} = \|\mathcal{E}_{\text{emo}}(\mathbf{Z}_0) - \mathbf{z}_e\|_2^2,
\end{equation}
where $\mathcal{E}_{\text{emo}}$ is the emotion encoder and $\mathbf{z}_e$ is the target emotion embedding. The total training objective is formulated as follows:
\begin{equation}
\mathcal{L}_{\text{total}} = \lambda_{\text{diff}}\mathcal{L}_{\text{diff}} + \lambda_{\text{sync}}\mathcal{L}_{\text{sync}} + \lambda_{\text{emo}}\mathcal{L}_{\text{emo}},
\end{equation}
where $\mathcal{L}_{\text{diff}}$ is the standard diffusion MSE loss, and $\{\lambda_{\text{diff}}, \lambda_{\text{sync}}, \lambda_{\text{emo}}\}$ are the corresponding weights.

\section{Experiments}

\begin{table}[t!]
  \centering
  \caption{Quantitative comparisons on MEAD-neu / HDTF / RAVDESS-neu datasets. \textbf{Bold} and \underline{underlined} notations denote Top-2 results.}
  \resizebox{\columnwidth}{!}{
    \begin{tabular}{lcccc}
      \toprule
      Methods & LSE-C $\uparrow$ & LMD $\downarrow$ & Blink $\uparrow$ & AU-std $\uparrow$ \\
      \midrule
      GT-FLAME & 4.29/4.82/3.41 & 0.00/0.00/0.00 & 0.15/0.26/0.37 & 0.15/0.17/0.20 \\
      MeshTalk~\cite{meshtalk} & 1.30/1.54/1.22 & 9.86/8.93/11.04 & 0.00/0.00/0.00 & 0.01/0.06/0.00 \\
      FaceFormer~\cite{faceformer} & 4.61/5.28/3.34 & \underline{5.93}/5.01/\underline{7.03} & 0.00/0.00/0.00 & 0.13/0.13/\underline{0.18} \\
      Imitator~\cite{imitator} & \textbf{5.64}/\textbf{6.41}/\textbf{4.07} & 7.26/6.39/8.51 & 0.00/0.00/0.00 & 0.13/0.12/0.17 \\
      Codetalker~\cite{codetalker} & 3.86/5.37/2.42 & 6.21/5.17/7.32 & 0.00/0.00/0.00 & 0.10/0.13/0.15 \\
      Emotalk~\cite{emotalk} & \underline{5.17}/5.24/3.31 & 6.08/\underline{4.92}/7.26 & 0.01/0.08/0.00 & \underline{0.18}/\textbf{0.21}/0.17 \\
      Emote~\cite{emote} & 3.52/4.02/2.12 & 9.92/8.94/8.91 & 0.00/0.00/0.00 & 0.00/0.00/0.00 \\
      FaceDiffuser~\cite{facediffuser} & 3.86/4.26/3.49 & 11.65/8.14/6.65 & \underline{0.09}/\underline{0.10}/\underline{0.28} & 0.12/0.11/0.14 \\

      \rowcolor[gray]{0.9}
      Ours & 5.05/\underline{6.04}/\underline{3.72} & \textbf{2.95}/\textbf{2.64}/\textbf{3.81} & \textbf{0.15}/\textbf{0.54}/\textbf{0.31} & \textbf{0.18}/\underline{0.20}/\textbf{0.22} \\
      \bottomrule
    \end{tabular}%
  }
  \label{tab:comparison_neutral}
\end{table}

\subsection{Implementation Details and Metrics}

\noindent\textbf{Datasets.} Our FLAME-centered multimodal emotion binding strategy is trained on the MEAD \cite{mead} and RAVDESS \cite{RAVDESS} datasets, while our attention-based latent diffusion model is trained on HDTF \cite{hdtf}, MEAD, and RAVDESS datasets. All datasets are processed through a FLAME-based 3D face reconstruction model \cite{emoca} to obtain FLAME expression parameters. We evaluate our approach on both neutral content (HDTF, MEAD-neu, and RAVDESS-neu) and emotional content (MEAD-emo and RAVDESS-emo).

\noindent\textbf{Implementation Details.} We employ the Adam optimizer for training. The multimodal binding strategy, comprising audio, FLAME, text, and label encoders, is trained on the MEAD and RAVDESS datasets to learn modality-invariant emotional features. The diffusion model is jointly trained on MEAD, RAVDESS, and HDTF datasets with 400 diffusion steps. The loss weights $\{\lambda_{\text{diff}}, \lambda_{\text{sync}}, \lambda_{\text{emo}}\}$ are set to 1.0, 0.01, and 0.01 respectively, balancing motion synthesis quality, speech synchronization, and emotional consistency. Training is conducted on a single NVIDIA A40 GPU for approximately 100k iterations with a learning rate of 0.0001. Detailed architecture specifications and parameter settings are provided in the appendix.

\noindent\textbf{Metrics.} To evaluate our framework's ability to address the two main challenges, we employ the following metrics: (1) For speech-motion synchronization, we use the lip synchronization error-confidence (LSE-C) \cite{syncnet} and landmark distance (LMD) of the mouth region \cite{atvg}. (2) For emotion control effectiveness, we utilize a SOTA image-based emotion classification model \cite{emo-sim} to compute emotion embedding similarity (Emo-sim). (3) To assess motion diversity and naturalness, we measure the blinking frequency and the standard deviation of facial action unit activation (AU-std) \cite{au}. Following physiological standards (0.28-0.45 blinks/s) \cite{blink}, we use blink rate (Blink) as an additional naturalness metric.

\subsection{Experimental Results}
\begin{table}[t!]
  \centering
  \caption{Quantitative comparisons on \textbf{MEAD-emo} / \textbf{RAVDESS-emo} datasets.}
  \resizebox{\columnwidth}{!}{
    \begin{tabular}{lccccc}
      \toprule
      Methods & LSE-C $\uparrow$ & LMD $\downarrow$ & Blink $\uparrow$ & AU-std $\uparrow$ & Emo-sim $\uparrow$ \\
      \midrule
      GT-FLAME & 3.83/3.01 & 0.00/0.00 & 0.21/0.30 & 0.15/0.17 & 1.00/1.00 \\
      MeshTalk~\cite{meshtalk} & 1.23/1.17 & 11.04/9.60 & 0.00/0.00 & 0.01/0.01 & 0.56/0.60 \\
      FaceFormer~\cite{faceformer} & 4.76/3.48 & \underline{7.03}/5.64 & 0.00/0.00 & 0.13/0.18 & 0.62/0.65 \\
      Imitator~\cite{imitator} & \textbf{5.96}/\textbf{4.35} & 8.51/7.33 & 0.00/0.00 & 0.13/0.16 & 0.52/0.60 \\
      Codetalker~\cite{codetalker} & 4.17/2.33 & 7.32/5.78 & 0.01/0.02 & 0.11/0.15 & 0.65/0.72 \\
      Emotalk~\cite{emotalk} & \underline{5.12}/3.43 & 7.26/\underline{5.50} & 0.01/0.15 & \underline{0.19}/\textbf{0.20} & 0.36/0.54 \\
      Emote~\cite{emote} & 3.24/1.86 & 10.35/8.62 & \underline{0.04}/0.16 & 0.00/0.00 & \underline{0.74}/\underline{0.79} \\
      FaceDiffuser~\cite{facediffuser} & 3.81/3.28 & 11.65/5.64 & \underline{0.28}/0.00 & 0.12/\underline{0.18} & 0.32/0.65 \\
      \rowcolor[gray]{0.9}
      Ours & 5.06/\underline{3.61} & \textbf{3.99}/\textbf{3.16} & \textbf{0.37}/\textbf{0.31} & \textbf{0.22}/\underline{0.18} & \textbf{0.92}/\textbf{0.94} \\
      \bottomrule
    \end{tabular}%
  }
  \label{tab:comparison_emo}
\end{table}
We compare our method with SOTA 3D face animation methods, including four emotion-agnostic methods: MeshTalk \cite{meshtalk}, FaceFormer \cite{faceformer}, Imitator \cite{imitator}, Codetalker \cite{codetalker}; two emotion-aware methods: Emotalk \cite{emotalk}, Emote \cite{emote}; and one diffusion-based method: FaceDiffuser \cite{facediffuser}. The reported results were obtained using the publicly available code of these methods.

\begin{figure*}[!ht]
  \centering
  \includegraphics[width=0.9\textwidth]{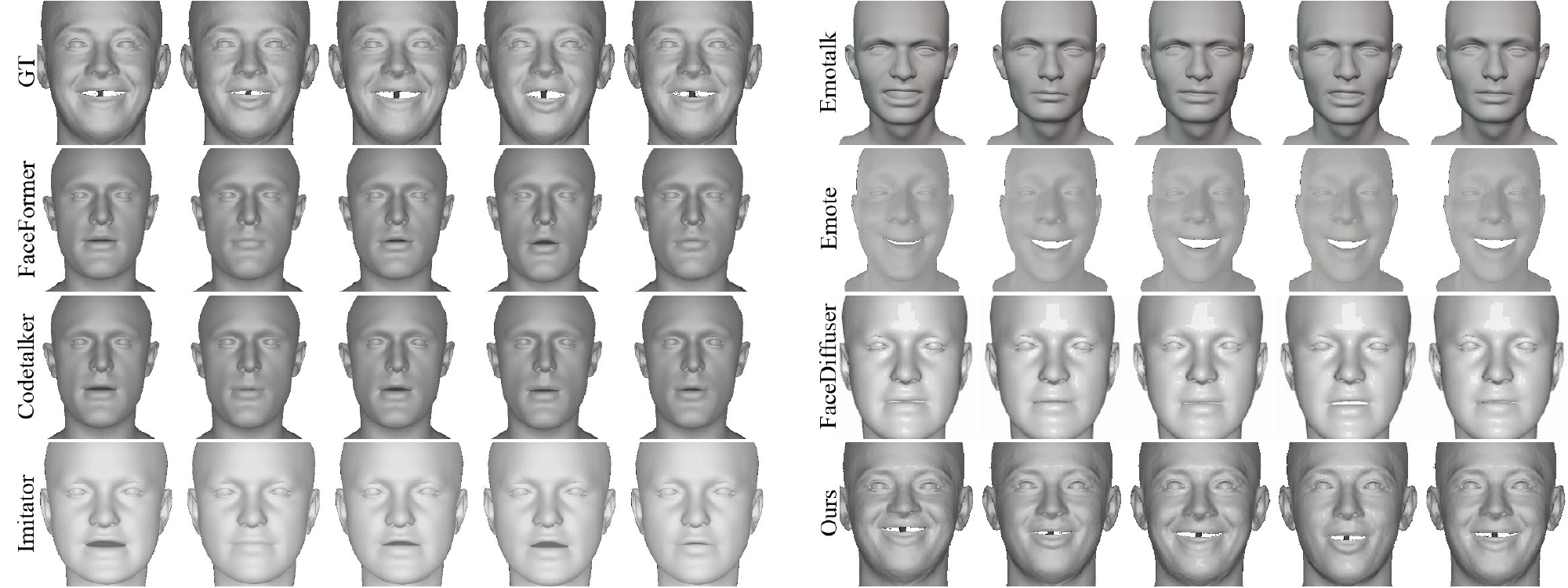}
  \caption{Qualitative comparison with SOTA methods on the MEAD-emo dataset. Our method generates more natural facial animations with both accurate lip synchronization and rich emotional expressions. For more results, please refer to the supplementary video.}
  \label{fig:comparison}
\end{figure*}

\noindent\textbf{Quantitative Results.} We conduct comprehensive evaluations against SOTA methods on both neutral and emotional datasets. For neutral speech animation (\text{Tab.} \ref{tab:comparison_neutral}), our diffusion-based framework demonstrates superior performance in motion synthesis, achieving a 48\% reduction in LMD while maintaining competitive LSE-C scores for lip synchronization. The increased AU-std values and natural blinking patterns indicate our model's ability to generate diverse facial dynamics, addressing the limitations of deterministic approaches. In emotional speech animation (\text{Tab.} \ref{tab:comparison_emo}), our FLAME-centered multimodal emotion binding strategy significantly enhances emotional expressiveness, with emotion similarity scores (Emo-sim: 0.92/0.94) exceeding the best baseline by 21.6\%, demonstrating the effective ability of our method to transfer emotional expression.

Overall, these results validate our framework's effectiveness in addressing both challenges: achieving flexible emotion manipulation through multimodal binding and generating diverse yet natural facial dynamics through the attention-based latent diffusion model.

\noindent\textbf{Qualitative Results.} \text{Fig.} \ref{fig:comparison} presents qualitative comparisons on the MEAD-emo dataset. Existing methods show notable limitations: EmoTalk struggles with emotional consistency due to the inherent ambiguity in audio-emotion mapping, while Emote's discrete label-based approach leads to limited emotional dynamics. Other methods, while achieving lip synchronization, show limited motion diversity due to their deterministic nature. In contrast, our results demonstrate both flexible emotion manipulation and diverse motion synthesis, with precise lip synchronization and natural emotional transitions.


\subsection{Ablation Study}

\begin{table}[t!]
  \centering
  \caption{Ablation studies on \textbf{MEAD-emo}. \textit{Input Modality}: Audio, Label, Text, Video; \textit{Diffusion Weight}: 0.5-3.0.}
  \resizebox{0.85\columnwidth}{!}{
    \begin{tabular}{lccccc}
      \toprule
      Methods & LSE-C $\uparrow$ & LMD $\downarrow$ & Blink $\uparrow$ & AU-std $\uparrow$ & Emo-sim $\uparrow$ \\
      \midrule
      Audio & 4.509 & 4.255 & 0.113 & 0.182 & 0.758 \\
      Label & 4.982 & 4.834 & 0.412 & 0.180 & 0.689 \\
      Text & 4.807 & 4.378 & 0.319 & 0.156 & 0.822 \\
      Video & 5.058 & 3.812 & 0.314 & 0.203 & 0.920 \\
      \midrule
      Weight = 0.5 & 2.305 & 3.892 & 0.246 & 0.117 & 0.937 \\
      Weight = 1.0 & 4.131 & 3.759 & 0.274 & 0.139 & 0.926 \\
      Weight = 1.5 & 4.894 & 3.767 & 0.322 & 0.172 & 0.920 \\
      Weight = 2.0 & 5.058 & 3.812 & 0.314 & 0.203 & 0.920 \\
      Weight = 3.0 & 5.060 & 3.992 & 0.371 & 0.224 & 0.922 \\
      \bottomrule
    \end{tabular}%
  }
  \label{tab:ablation_1}
\end{table}

\begin{figure}[!t]
  \centering
  \includegraphics[width=0.45\textwidth]{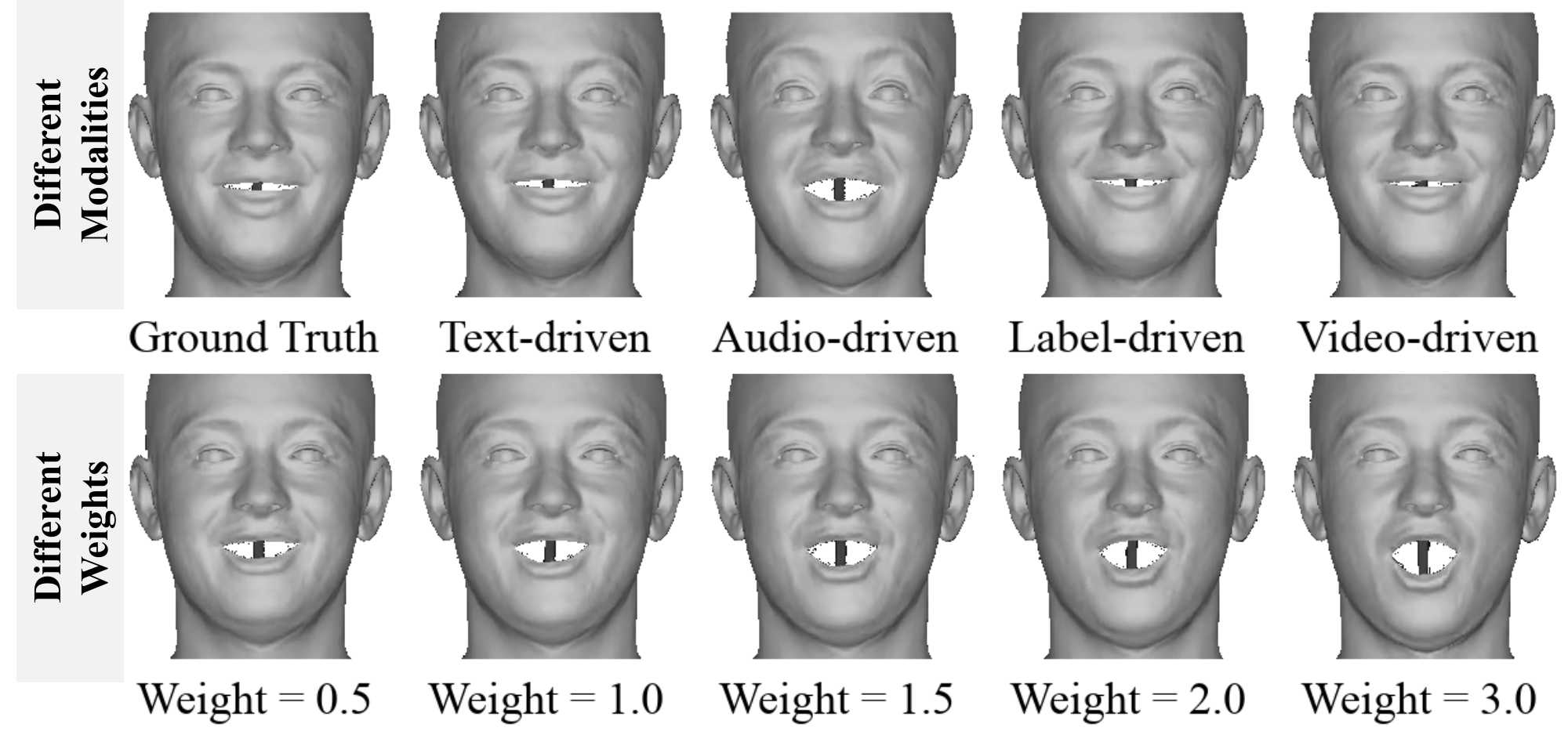}
  \caption{Qualitative results of our method. Top: Results with different emotional control modalities. Bottom: Results with varying diffusion weights, demonstrating the trade-off between motion diversity and emotional consistency.}
  \label{fig:ablation}
\end{figure}
We conduct comprehensive ablation experiments to validate our key technical contributions.

\noindent\textbf{FLAME-Centered Multimodal Emotion Binding Strategy.} Our multimodal binding strategy demonstrates effective emotion control across different input modalities (\text{Tab.} \ref{tab:ablation_1}, top). Video-based control achieves superior performance (Emo-sim: 0.920, LMD: 3.812), while text modality shows competitive results (Emo-sim: 0.822). As illustrated in \text{Fig.} \ref{fig:ablation}, our method maintains consistent emotional expression quality across all modalities, validating the effectiveness of our FLAME-centered binding mechanism.

\noindent\textbf{Style Weight Analysis.} We analyze the impact of style weight scaling using video-based emotional prompts (\text{Tab.} \ref{tab:ablation_1} and Fig. \ref{fig:ablation}, bottom). A weight of 2.0 achieves the optimal balance between emotion transfer and geometric accuracy (LSE-C: 5.058, AU-std: 0.203).


\begin{table}[t!]
  \centering
  \caption{Ablation study on different components.}
  \resizebox{1\columnwidth}{!}{
    \begin{tabular}{lcccc}
      \toprule
      Methods & LSE-C $\uparrow$ & Blink $\uparrow$ & AU-std $\uparrow$ & Emo-sim $\uparrow$ \\
      \midrule
      \textsc{Former} & 0.445 & 0.101 & 0.123 & 0.880 \\
      \textsc{Former}+\textsc{sync} & 3.865 & 0.002 & 0.030 & 0.824 \\
      \textsc{Former}+\textsc{sync}+\textsc{emo} & \underline{4.523} & 0.146 & \underline{0.135} & \underline{0.916} \\
      Ours w/o \textsc{sync} w/o \textsc{emo} & 1.363 & \textbf{0.331} & 0.123 & 0.848 \\
      Ours w/o \textsc{align mask} & 1.383 & \textbf{0.338} & 0.121 & 0.867 \\
      \rowcolor[gray]{0.9}
      Ours Full & \textbf{5.058} & \textbf{0.314} & \textbf{0.203} & \textbf{0.920} \\
      \bottomrule
    \end{tabular}%
  }
  \label{tab:ablation_2}%
\end{table}%

\noindent\textbf{Architecture Components.} The baseline deterministic transformer (\textsc{Former}) exhibits the limitations discussed earlier, showing restricted performance across all metrics. Adding synchronization loss (\textsc{Former}+\textsc{sync}) improves lip sync accuracy (LSE-C: 3.865) but, due to its deterministic nature, significantly compromises motion diversity (AU-std: 0.030). Further incorporating emotion loss (\textsc{Former}+\textsc{sync}+\textsc{emo}) helps maintain emotional consistency while achieving synchronization, but still suffers from limited motion diversity. Our full model, combining the attention-based latent diffusion architecture with FLAME-centered multimodal emotion binding strategy, successfully addresses both challenges, achieving superior performance in both emotion control (Emo-sim: 0.920) and motion diversity (AU-std: 0.203). The ablation studies of removing synchronization and emotion losses (w/o \textsc{sync} w/o \textsc{emo}) or alignment mask (w/o \textsc{align mask}) further validate that each component plays a crucial role in achieving these dual objectives.

\noindent\textbf{\textit{Question: Can our multimodal binding strategy be applied to deterministic-based 3D face animation?}}

To investigate this question, we integrate our binding strategy with deterministic-based methods. As shown in \text{Tab.} \ref{tab:ablation_2}, while achieving comparable emotion similarity scores, deterministic methods exhibit significantly lower AU-std values compared to our diffusion-based approach. This validates both the effectiveness of our multimodal binding strategy and the necessity of our attention-based latent diffusion model for generating diverse facial dynamics.


\section{Conclusion}

In this paper, we present a diffusion-based framework for expressive 3D facial animation that effectively addresses two fundamental challenges in existing methods: single-modal emotional control and deterministic motion generation. Our framework introduces two key technical innovations: a FLAME-centered multimodal emotion binding strategy that unifies emotional representations from diverse modalities through contrastive learning, and an attention-based latent diffusion model that captures the stochastic nature of emotional expressions while preserving speech synchronization. Extensive experiments demonstrate the superiority of our approach across different input modalities, achieving significant improvements in both emotion similarity and motion diversity while maintaining precise lip synchronization and emotional coherence. We believe our work provides a promising direction for generating more expressive and natural 3D facial animations, benefiting various applications in virtual reality and digital entertainment.


\bibliographystyle{IEEEbib}
\bibliography{icme2025references}

\begin{thebibliography}{10}

\bibitem{vocaset}
Cudeiro D, Bolkart T, and Laidlaw C,
\newblock ``Capture, learning, and synthesis of 3d speaking styles,''
\newblock in {\em CVPR}, 2019.

\bibitem{meshtalk}
Richard A, Zollhöfer M, Wen Y, et~al.,
\newblock ``Meshtalk: 3d face animation from speech using cross-modality disentanglement,''
\newblock in {\em ICCV}, 2021, pp. 1173--1182.

\bibitem{faceformer}
Yingruo Fan, Zhaojiang Lin, Jun Saito, Wenping Wang, and Taku Komura,
\newblock ``Faceformer: Speech-driven 3d facial animation with transformers,''
\newblock in {\em CVPR}, 2022.

\bibitem{codetalker}
Xing J, Xia M, and Zhang Y,
\newblock ``Codetalker: Speech-driven 3d facial animation with discrete motion prior,''
\newblock in {\em CVPR}, 2023.

\bibitem{emote}
Radek Dan{\v{e}}{\v{c}}ek, Chhatre, et~al.,
\newblock ``Emotional speech-driven animation with content-emotion disentanglement,''
\newblock in {\em SIGGRAPH Asia}, 2023, pp. 1--13.

\bibitem{emotalk}
Ziqiao Peng, Haoyu Wu, et~al.,
\newblock ``Emotalk: Speech-driven emotional disentanglement for 3d face animation,''
\newblock in {\em ICCV}, 2023.

\bibitem{imitator}
Thambiraja B, Habibie I, and Aliakbarian S,
\newblock ``Imitator: Personalized speech-driven 3d facial animation,''
\newblock in {\em ICCV}, 2023.

\bibitem{mead}
Kaisiyuan Wang, Qianyi Wu, Linsen Song, et~al.,
\newblock ``Mead: A large-scale audio-visual dataset for emotional talking-face generation,''
\newblock in {\em ECCV}, 2020.

\bibitem{styletalk}
Yifeng Ma, Suzhen Wang, Zhipeng Hu, Changjie Fan, Tangjie Lv, Yu~Ding, Zhidong Deng, and Xin Yu,
\newblock ``Styletalk: One-shot talking head generation with controllable speaking styles,''
\newblock {\em CoRR}, 2023.

\bibitem{flame}
Li~T, Bolkart T, and Black~M J,
\newblock ``Learning a model of facial shape and expression from 4d scans,''
\newblock {\em ACM Trans. Graph.}, vol. 36, no. 6, 2017.

\bibitem{avsp}
Michael~M. Cohen, Rashid Clark, and Dominic~W. Massaro,
\newblock ``Animated speech: research progress and applications,''
\newblock in {\em {AVSP}}, 2001, p. 200.

\bibitem{wav2vec2}
Alexei Baevski, Yuhao Zhou, Abdelrahman Mohamed, and Michael Auli,
\newblock ``wav2vec 2.0: {A} framework for self-supervised learning of speech representations,''
\newblock in {\em NIPS}, 2020.

\bibitem{hubert}
Wei{-}Ning Hsu, Benjamin Bolte, et~al.,
\newblock ``Hubert: Self-supervised speech representation learning by masked prediction of hidden units,''
\newblock {\em TASLP}, 2021.

\bibitem{sadtalker}
Wenxuan Zhang, Xiaodong Cun, et~al.,
\newblock ``Sadtalker: Learning realistic 3d motion coefficients for stylized audio-driven single image talking face animation,''
\newblock {\em CVPR}, 2023.

\bibitem{talkface}
Zhang C, Ni~S, and Fan Z,
\newblock ``3d talking face with personalized pose dynamics,''
\newblock {\em TVCG}, vol. 29, no. 2, pp. 1438--1449, 2021.

\bibitem{produb}
Kangwei Liu, Xiaowei Yi, and Xianfeng Zhao,
\newblock ``Produb: Progressive growing of facial dubbing networks for enhanced lip sync and fidelity,''
\newblock in {\em ICME}. IEEE, 2024.

\bibitem{facediffuser}
Stan S, Haque~K I, and Yumak Z,
\newblock ``Facediffuser: Speech-driven 3d facial animation synthesis using diffusion,''
\newblock in {\em SIGGRAPH Conference on Motion, Interaction and Games}, 2023, pp. 1--11.

\bibitem{diffusionmotion}
Zhiyuan Ren, Zhihong Pan, Xin Zhou, and Le~Kang,
\newblock ``Diffusion motion: Generate text-guided 3d human motion by diffusion model,''
\newblock in {\em ICASSP}. IEEE, 2023, pp. 1--5.

\bibitem{biwi}
Ruiz N, Chong E, and Rehg~J M,
\newblock ``Fine-grained head pose estimation without keypoints,''
\newblock in {\em CVPR workshops}, 2018.

\bibitem{multiface}
Wuu C, Zheng N, and Ardisson S,
\newblock ``Multiface: A dataset for neural face rendering,''
\newblock {\em arXiv}, 2022.

\bibitem{RAVDESS}
Steven~R Livingstone and Frank~A Russo,
\newblock ``The ryerson audio-visual database of emotional speech and song (ravdess): A dynamic, multimodal set of facial and vocal expressions in north american english,''
\newblock {\em PloS one}, vol. 13, no. 5, pp. e0196391, 2018.

\bibitem{emoca}
Daněček R, Black~M J, and Bolkart T,
\newblock ``Emoca: Emotion driven monocular face capture and animation,''
\newblock in {\em CVPR}, 2022.

\bibitem{attention}
Ashish Vaswani, Noam Shazeer, et~al.,
\newblock ``Attention is all you need,''
\newblock in {\em NIPS}, 2017.

\bibitem{roberta}
Liu Y,
\newblock ``Roberta: A robustly optimized bert pretraining approach,''
\newblock {\em arXiv}, vol. 364, 2019.

\bibitem{seeg}
Liang Y, Feng Q, Zhu L, et~al.,
\newblock ``Seeg: Semantic energized co-speech gesture generation,''
\newblock in {\em CVPR}, 2022, pp. 10473--10482.

\bibitem{film}
Ethan Perez, Florian Strub, Harm de~Vries, Vincent Dumoulin, and Aaron~C. Courville,
\newblock ``Film: Visual reasoning with a general conditioning layer,''
\newblock in {\em AAAI}, 2018, pp. 3942--3951.

\bibitem{adain}
Xun Huang and Serge~J. Belongie,
\newblock ``Arbitrary style transfer in real-time with adaptive instance normalization,''
\newblock in {\em ICCV}, 2017.

\bibitem{classfree}
Jonathan Ho and Tim Salimans,
\newblock ``Classifier-free diffusion guidance,''
\newblock {\em arXiv}, 2022.

\bibitem{syncnet}
Joon~Son Chung and Andrew Zisserman,
\newblock ``Out of time: Automated lip sync in the wild,''
\newblock in {\em ACCV}, 2016.

\bibitem{hdtf}
Zhimeng Zhang, Lincheng Li, Yu~Ding, and Changjie Fan,
\newblock ``Flow-guided one-shot talking face generation with a high-resolution audio-visual dataset,''
\newblock in {\em CVPR}, 2021.

\bibitem{atvg}
Lele Chen, Ross~K. Maddox, Zhiyao Duan, and Chenliang Xu,
\newblock ``Hierarchical cross-modal talking face generation with dynamic pixel-wise loss,''
\newblock in {\em CVPR}, 2019.

\bibitem{emo-sim}
Junjie Li, Jin Yuan, and Zhiyong Li,
\newblock ``{TP-FER:} an effective three-phase noise-tolerant recognizer for facial expression recognition,''
\newblock {\em {ACM} Trans. Multim. Comput. Commun. Appl.}, 2023.

\bibitem{au}
P~Ekman and W~V Friesen,
\newblock ``Facial action coding system,''
\newblock {\em Environmental Psychology and Nonverbal Behavior}, 1978.

\bibitem{blink}
Sanjana Sinha, Sandika Biswas, and Brojeshwar Bhowmick,
\newblock ``Identity-preserving realistic talking face generation,''
\newblock in {\em {IJCNN}}. 2020, pp. 1--10, {IEEE}.

\end{thebibliography}


\end{document}